\documentclass[proof]{pasj01}
\draft

\def\kms{km s$^{-1}$}
\def\h2o{H$_2$O}
\def\nh3{NH$_3$}
\def\vlsr{$v_{\rm LSR}$}
\def\hii{H{\rm II}}
\def\afg{AFGL\,333-Ridge}

\newcommand{\red}{\textcolor{red}}

\usepackage{natbib}
\usepackage{color}

\begin{document}
\Received{2016/07/07}
\Accepted{2016/11/18}

\title{Interaction Between  \hii\ Region and \afg: Implications to the Star Formation Scenario}

\author{
Makoto \textsc{Nakano}\altaffilmark{1.*}, 
Takashi \textsc{Soejima}\altaffilmark{1}, 
James O. \textsc{Chibueze}\altaffilmark{2}
Takumi \textsc{Nagayama}\altaffilmark{3},
Toshihiro \textsc{Omodaka}\altaffilmark{4}, 
Toshihiro \textsc{Handa}\altaffilmark{4}, 
Kazuyoshi \textsc{Sunada}\altaffilmark{3}, 
Tatsuya \textsc{Kamezaki}\altaffilmark{4}, 
\and
Ross A. \textsc{Burns}\altaffilmark{4,5}}

\altaffiltext{1}{Faculty of Education, Oita University, 700 Dannoharu, Oita 870-1192, Japan.}
\altaffiltext{2}{Department of Physics and Astronomy, Faculty of Physical Sciences,\\
	 University of Nigeria, Carver Building, 1 University Road, Nsukka, Nigeria.}
 \altaffiltext{3}{Mizusawa VLBI Observatory, National Astronomical Observatory of Japan, 2-21-1 Osawa, Mitaka, Tokyo 181-8588, Japan.}
 \altaffiltext{4}{Department of Physics and Astronomy, Graduate School of Science and Engineering, Kagoshima University, 1-21-35 Korimoto, Kagoshima 890-0065, Japan.}
 \altaffiltext{5}{Joint Institute for VLBI ERIC (JIVE), Postbus 2, 7990 AA Dwingeloo, the Netherlands}

\email{mnakano@oita-u.ac.jp}

\maketitle

\begin{abstract}
We investigated the star formation activities in the AFGL\,333 region, which 
is in the vicinity of the W4 expanding bubble, by conducting 
\nh3\ (1,1), (2,2), and (3,3) mapping observations with the 45 m 
Nobeyama Radio Telescope at an angular resolution of 75\arcsec. 
The morphology of the \nh3\ (1,1) map shows a bow-shape structure
with the size of 2.0 $\times$ 0.6 pc 
as seen in the dust continuum. 
At the interface between the W4 bubble and the dense \nh3\ cloud,  
the compact \hii\ region G\,134.2+0.8, associated with IRAS\,02245$+$6115, is located.
Interestingly, just north and south of G\,134.2+0.8 we found \nh3\ emission exhibiting large velocity widths 
of $\sim 2.8$ \kms, compared to 1.8 \kms at the other positions.  
As the possibility of mechanical energy injection through the activity of YSO(s)  is low,
we considered the origin of the large turbulent gas motion as indication of interaction between the 
compact \hii\ region and the periphery of the dense molecular cloud. 
We also found expanding motion of the CO 
emission
associated with G\,134.2+0.8.
The 
overall
structure of the AFGL\,333-Ridge might have been formed by the expanding
bubble of W4. 
However, the small velocity widths observed west of
IRAS\,02245$+$6115, around the center of the dense molecular cloud, 
suggest that interaction with the compact \hii\ region is limited.
Therefore the YSOs (dominantly Class 0/I) in the core of the \afg\ dense molecular cloud  most 
likely formed in quiescent mode. 
As has been previously suggested for the large scale star formation in the W3 giant molecular cloud, 
our results 
show an apparent coexistence of induced and quiescent star formation in this 
region. It appears that star formation in the AFGL\,333 region has proceeded without significant external triggers, 
but accompanying stellar feedback environment.
\end{abstract}

\section{Introduction}

The influence of \hii\ regions on their immediate surrounding, 
especially in inducing star formation activities, has drawn a 
lot of attentions in recent times.
\cite{ell77} first proposed that high-mass stars are 
formed through a 
``collect-and-collapse'' sequential process 
triggered by the influence of an expanding \hii\ region. 
Later, \cite{whi94}  explained why gravitationally unstable fragments 
are formed in the shocked layers which fragment to produce high-mass stars.  
Since then many observational studies have targeted triggered star formation 
\citep[e.g.][]{car00,lef00,deh05}. 
Although  examples of quiescent formation of low mass stars have been reported in regions such as
Taurus, those for high mass stars are rare, KR\,140 \citep{bal00} being one example. 
While the influence of expanding \hii\ regions on their surroundings 
cannot be denied, \cite{chi13} argued an isolated 
case of the star associated with S\,252A forming spontaneously 
though located near the shock fronts of the Monkey Head Nebula 
(Gem OB1). Identifying the formation scenario 
requires studying 
the cloud properties and star formation in regions under the influence 
of \hii\ regions. 

W3\,Main, W3\,(OH) and AFGL\,333 are active massive star forming regions 
located in the high-density layer (HDL) on the western edge of the
W4 expanding \hii\ region, or bubble \citep{lad78}. 
W4 is excited by a group of O stars located at the central part of the \hii\ region \citep{lef97}. 
\cite{oey05} suggested  a three generation system 
of hierarchical triggered star formation, based on the stellar populations of IC\,1795. 
The first generation is the 1300 pc loop and the Perseus superbubble. 
These triggered the formation of IC\,1795, and then IC\,1795 triggered the formation of 
W3\,North, W3\,Main, W3\,OH at the edge of shell. An X-ray study by \cite{fei08} 
revealed the diversity of the W3 stellar populations. 
\cite{roc11} analyzed the protoplanetary disks of young stars in the IC\,1795 region as 
massive star forming region by a deep Spitzer survey.
The wider area including AFGL\,333 and KR\,140 has been investigated with Spitzer and Herschel \citep{riv11,riv13}.
Observational results of a sub-mm survey of dense clumps by \cite{moo07},
and extensive molecular line observations by \cite{pol12}
agree well with the  collect-and-collapse model. 
\cite{bie11} also produced the extensive CO maps of the W3 region.
Star formation in W3\,Main and W3\,(OH) are considered to have been triggered by IC\,1795. 
On the other hand KR\,140, which is located in the
southwest quadrant of W3 giant molecular cloud (GMC), appears to be formed in quiescently.
Recently much of works have been done to reveal the star formation history of the W3 complex.
The stellar contents in W3\,Main \citep{bik14}, IC\,1795 \citep{rom15}, 
and across the complex \citep{kim15} were investigated. 
The statistical study was done by using Herschel datasets \citep{riv15}. 
These studies showed a complex history of star formation in the W3 GMC,
and suggested  massive stars and clusters in the active region as
W3\,Main and W3\,(OH) were formed in the dense material by the progression 
of star formation process.  Although triggering events play an 
important role to form the active region, it is still unclear in more 
quiescent region as AFGL\,333 or KR\,140.
\afg\ is a dense molecular cloud located south of W3(OH). 
It is associated with three known IRAS sources, namely, 
IRAS\,02252$+$6120 
\citep[SFO\,05:][]{sug91}, IRAS\,02244$+$6117, and IRAS\,02245$+$6115 (AFGL\,333)
(See figure \ref{fig:01}).
Although AFGL\,333 is located in the HDL, its star formation is less active than the other two
regions (W3\,Main and W3\,(OH))  and
it has less extreme environmental conditions  \citep[e.g.][]{riv13}.
The deep NIR photometry by \cite{jos16} unraveled the star formation activity of AFGL\,333 region. 
The AFGL\,333 region is one of the best sites to
study the influence of 
\hii\ regions  to star formation because of 
its location; the dense molecular cloud at the edge of the extended bubble (Heart Nebula) 
of W4 and the presence of a  compact \hii\ region in 
IRAS\,02245$+$6115.

To study the interaction between  \hii\ regions and 
molecular clouds, investigation of the distribution of 
dense gas and of the gas temperature is vital. \nh3\ lines 
prove to be particularly useful for investigating the physical conditions of 
dense gas clumps of a molecular cloud. 
The \nh3\ observations have advantages compared to other lines, such as $^{13}$CO and C$^{18}$O, 
because \nh3\ is an intermediate-density gas tracer, and gives gas temperature using the $(2,2)/(1,1)$ line ratio.
The information on gas motion and temperature distribution can indicate signs of interaction.  
It also reveals the presence of interstellar shocks through the $(3,3)/(1,1)$ line ratio, shown by \cite{nag09}. 
As example of such a study is given in \cite{urq11} who surveyed
massive young stellar objects and ultra-compact HII regions in 
ammonia -- allowing them to derive the basic physical properties of 
the associated high density gas for their sample. 
As such, we made 
mapping observations of the AFGL\,333 region using \nh3\ (1,1), 
(2,2), and (3,3) lines and estimated the distribution of the 
gas kinetic temperature over the cloud.
We also simultaneously observed \h2o\ maser emission which is a
well known tracer of early star formation activity 
\citep[e.g.][]{sun07}.

In this paper, we observationally investigate the 
dense clumps of the molecular cloud in the AFGL\,333 region that are influenced by the expanding \hii\ region, and 
compared our results with archival data/catalogs to reveal the 
local star formation scenario in this region.
The parallax distance of W3\,(OH) was obtained to be  2.0 kpc 
\citep{xu06,hac06}. Since the line of sight velocities of W4 
and W3\,(OH) are within of 5 \kms\ of each other,
we will assume a common distance of 2.0 kpc for the AFGL\,333 region in this paper.

\section{Observations}

\subsection{\nh3\ and \h2o\ Maser Observations}

\begin{figure}
\begin{center}
\includegraphics[width=15cm]{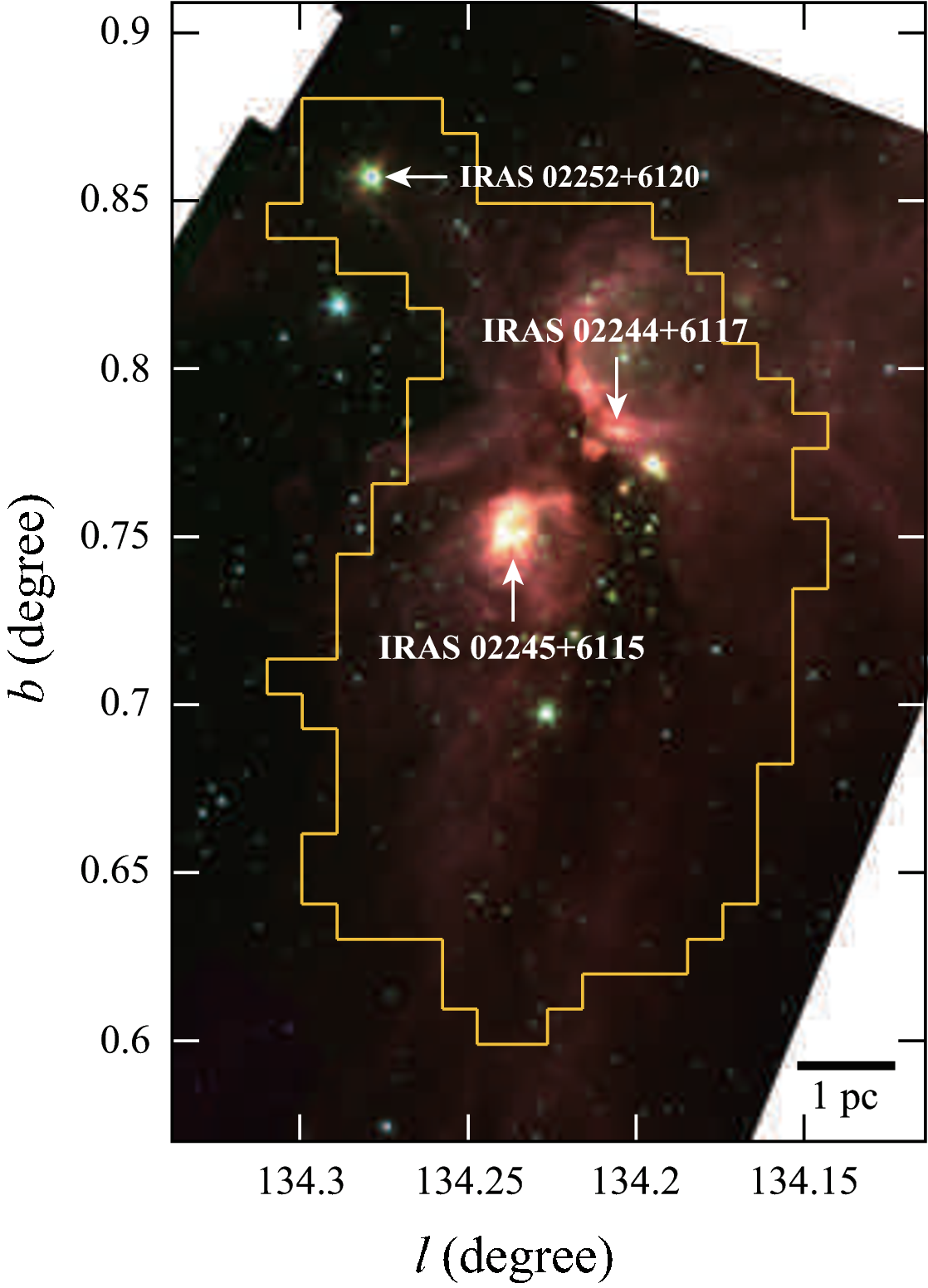}
\end{center}
\caption{Three-color (red: 8.0 $\mu$m, green: 5.8 $\mu$m, blue: 4.5 $\mu$m) Spitzer-IRAC image of the AFGL\,333 region.
         Yellow enclosure shows the mapping area of our \nh3\ observations.}
\label{fig:01}
\end{figure}

We conducted \nh3\ and \h2o\ maser observations of the AFGL\,333 region using 
the 45 m telescope at Nobeyama Radio Observatory (NRO) between 
January and June 2013. 
\nh3\ (J,K) = (1,1), (2,2), and (3,3) lines, 
whose frequencies are 23.694495 GHz, 23.722633 GHz, and 23.870129 GHz, 
respectively, and the 22.235080 GHz \h2o maser line were observed
simultaneously. 
We used the H22 receiver, a cooled HEMT receiver, 
and the SAM45, a digital spectrometer, with eight IFs
to observe both polarizations for each line simultaneously.
The bandwidth and spectral resolution of each IF were 62 MHz 
and 30.5 kHz, respectively 
corresponding to $\sim 800$ \kms\ and $\sim 0.39$ \kms\ for the 
observed lines. 
The telescope beam size at 23 GHz is 75\arcsec, which corresponds to 0.73 pc at 2.0 kpc. 
We observed 267 positions with a 37.5\arcsec\ grid 
along the Galactic coordinates using position switching between 
the target and the reference position comprising of the empty sky.
We investigated the kinematic and density structure of the 
AFGL\,333 region at clump scales, or 0.5-1.0 pc.
The reference position was taken at 
$(l, b) = (\timeform{134.3400D}, \timeform{0.5500D})$, 
where no \nh3\ or \h2o\ maser was detected. 
The target mapping area of  the AFGL\,333 region was $600\arcsec \times 1000\arcsec$ 
centered at $(l, b) = (\timeform{134.2000D}, \timeform{0.7500D})$.
Figure \ref{fig:01} shows the mapping area of the $l$-$b$ observations
overlaid on the Spitzer-IRAC image.
The pointing accuracy was checked using an \h2o\ maser source 
Cepheus A and was better than 5\arcsec.

Data reduction was carried out 
using the NEWSTAR software 
package developed by 
the Nobeyama Radio Observatory (NRO).
Emission free channels were used to estimate and subtract 
a spectral baseline.
We applied only the linear line as the continuum level.
Vertical and horizontal
polarizations were combined and weighted based on 
the system noise temperature which was between 100 and 300 K. 
The antenna temperature in Kelvin scale was measured using 
the chopper wheel method \citep{kut81}.
The median $1\sigma$ rms noise level was 0.04 K.
For the \h2o\ maser data. 
We converted the antenna temperature to the flux density 
using the conversion factor of 2.6 Jy K$^{-1}$.
In this paper, the intensities of \nh3\ and \h2o\ maser
are presented 
as the antenna temperature in  Kelvin and 
flux density in Jansky, respectively.

\subsection{Archival Data/Catalog}

We used  the data cubes of the J=2-1 lines of $^{12}$CO and $^{13}$CO obtained by
the Heinrich Hertz Submillimeter Telescope \citep{bie11}.
The effective resolution of the maps is 38\arcsec\ FWHM and the sampling velocity of 0.5 \kms.
1420 MHz radio continuum data from the 
Canadian Galactic Plane Survey  \citep[CGPS:][]{tay03}
was also used to analyse the ionized gas distribution.
The image was obtained with a synthesized beam and an rms noise of
1.0 $\times$ 0.81 arcmin$^2$ and 0.063 K.

To establish the distributions of young stars and dust, we
obtained Spitzer Space Telescope 3.6, 4.5, 5.8 and 8.0 $\mu$m  
IRAC image of the AFGL\,333 region 
from the Spitzer Heritage Archive 
hosted by the Infrared Science Archive at IPAC (Post-Basic Calibrated Data: Program ID 30995), 
which shows the bright IRAS sources 
(clusters) and a prominent infrared dark filament representing 
the \afg\ dense molecular cloud.
We also used the Spitzer YSO catalog in this region (Catalog 1 by  \cite{riv11}),
and the photometric data of all point sources within AFGL\,333 by \cite{jos16}.

\section{Results}

\subsection{Spatial Distribution of \nh3}

\begin{figure}
\begin{center}
\includegraphics[width=8cm]{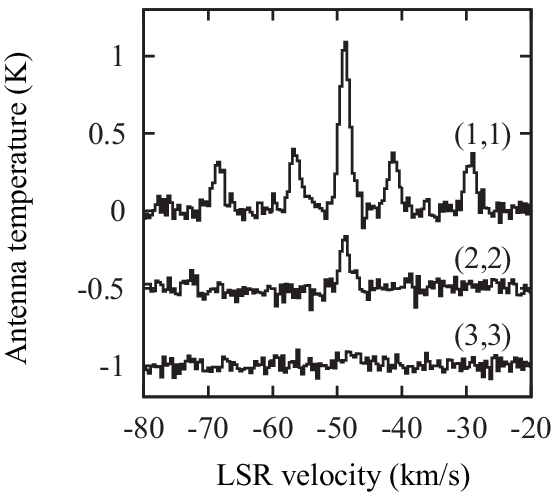}
\end{center}
\caption{\nh3 (1,1), (2,2) and (3,3) spectra towards the (1,1) emission peak
         at $(l, b) = (\timeform{134.2104D}, \timeform{0.7188D})$, shown in figure 3.
         }
\label{fig:02}
\end{figure}

\nh3\ $(J,K) =$ (1,1), (2,2), and (3,3) emission for which
the integrated intensities exceed the 2$\sigma$ noise level 
($\geq 0.14$ K \kms), were obtained
at 165, 111, and 71 positions out of the 267 observed positions, 
respectively.
Figure \ref{fig:02} shows the \nh3\ (1,1), (2,2), and (3,3) 
profiles towards the (1,1) emission peak at 
$(l, b) = (\timeform{134.2104D}, \timeform{0.7188D})$.
The \nh3\ line profile comprises five quadruple hyperfine 
components consisting of a main line and two symmetrical 
pairs of inner and outer satellite lines.
We could detect the inner and outer satellite lines of the (1,1) 
line in 51 positions. 
These 51 positions show (2,2) main lines with $> 2\sigma$,
and 36 points show (3,3) main lines among them.
Detections of both main and satellite lines in (1,1) 
lines yield an opacity estimation using their intensity 
ratio (see subsection 3.3).
The satellite lines of (2,2) and (3,3) line were not detected 
at any positions in the observed area.

\begin{figure}
\begin{center}
\includegraphics[width=12cm]{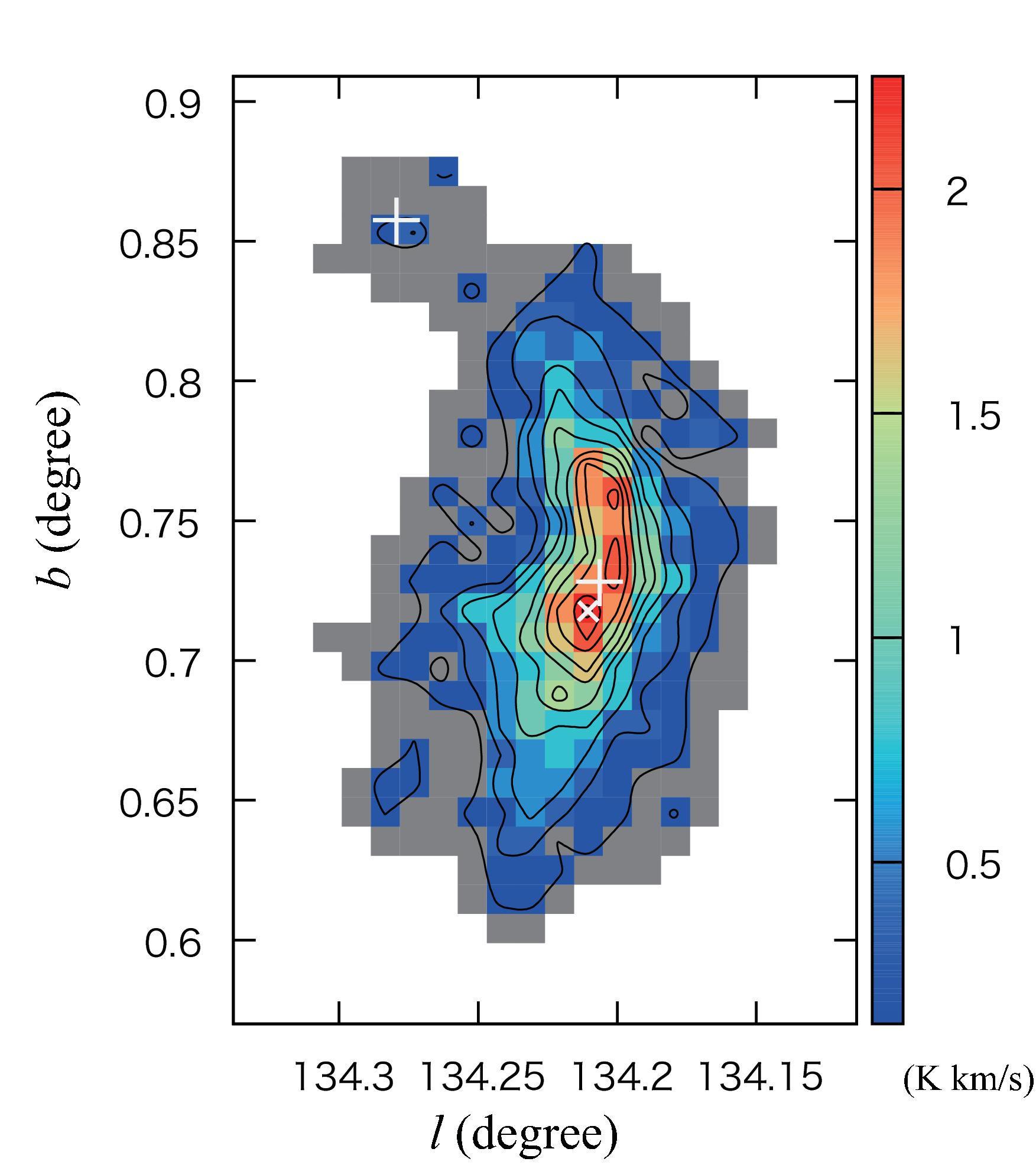}
\end{center}
\caption{Integrated intensity map of  \nh3\ (1,1) line.
         The lowest contour and contour interval are 
         0.14 K \kms\ (2$\sigma$) and 
         0.22 K \kms\, respectively.
         The integrated velocity range is from 
         $-52.0$ to $-45.5$ \kms.
         Pluses and cross denote the positions of two \h2o\ masers,
         and the (1,1) emission peak, respectively.}
\label{fig:03}
\end{figure}

Figure \ref{fig:03} shows the $l$-$b$ map of (1,1) line,
velocity-integrated in the range from \vlsr $= -52.0$ to
$-45.5$ \kms.
No significant (1,1) emission was detected beyond this 
velocity range.
(1,1) emission is distributed with an area of 
$\Delta l \times \Delta b$ ${\simeq}$ $\timeform{0.15D} \times \timeform{0.25D}$ 
(5.2 $\times$ 8.7 pc at the distance of 2.0 kpc). 
The overall morphology shows bow shape with the size of $3.{\arcmin}4 \times 1.{\arcmin}0$ ($2.0 \times 0.6$ pc)
at the level of half of peak intensity.
There is weak and compact emission at north-east of the ridge 
near $(l, b) \simeq (\timeform{134.27D}, \timeform{0.86D})$ 
corresponding to the position of SFO\,05 \citep{fuk13} at the northern tip
of the infrared dark filament.
SFO\,05 is a small bright-rimmed cloud with an optical size of $<$ 1\arcmin.
This bright-rimmed cloud was observed 
with the Green Bank telescope in \nh3\ (1,1) and (2,2) lines 
\citep{mor10}.
The bow-shaped distribution of the (1,1) line is similar to those of 
the $^{13}$CO and C$^{18}$O \citep{sak06}, submillimter observations
\citep{dif08}, Herschel far infrared map \citep{riv13}, and the filamentary 
structure of the dark infrared cloud \citep{riv11}.

\subsection{Velocity Distribution of \nh3}

\begin{figure}
\begin{center}
\includegraphics[width=15cm]{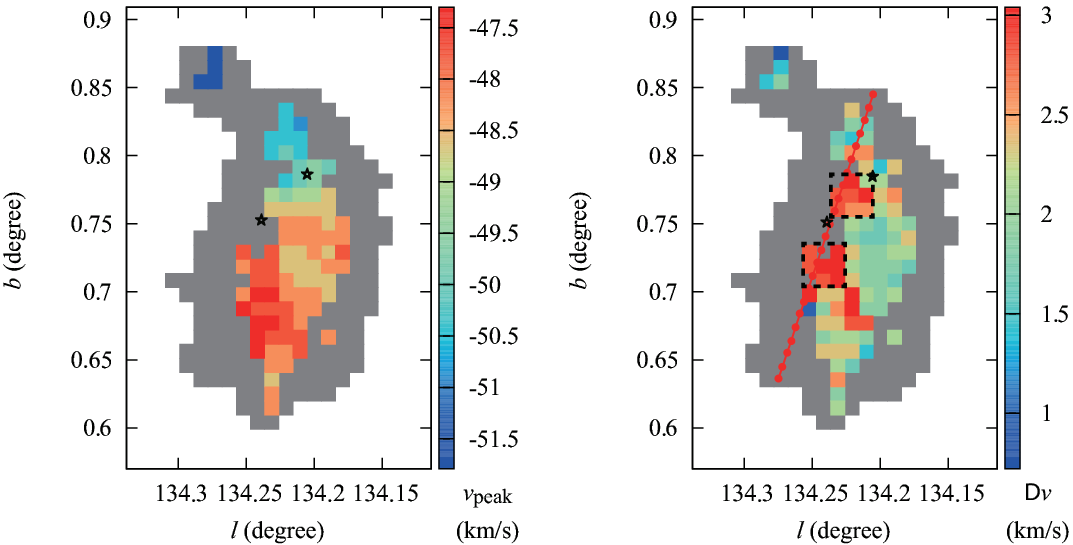}
\end{center}
\caption{Peak velocity (left) and velocity width (right) maps of \nh3\ (1,1) line.
	The region of large velocity width is shown by the dashed boxes. 
	Stars indicate the positions of IRAS\,02245$+$6115 and IRAS\,02244$+$6117.}
\label{fig:04}
\end{figure}

Figure \ref{fig:04} shows the peak velocity and velocity 
width maps of (1,1) line. 
The peak velocity and the velocity width were
obtained from the Gaussian fitting to the 
main line of the (1,1) spectrum.
We observe a north-south velocity gradient along 
the ridge from the peak velocity map.
The most blueshifted component of \vlsr $= -51.5$ \kms\
is detected near the position of SFO\,05.
This velocity is consistent with that of \cite{mor10}.
SFO\,05 should be on the near side of W4.
\cite{bie11} reported that the AFGL\,333 region shows the most negative 
velocity emission in the high density layer (HDL).
The velocity structure in the AFGL\,333 region is also consistent with the C$^{18}$O results by \cite{sak07},
blueshifted at the northside of the NH$_{3}$ peak, and redshifted at the southern side.
This may indicates that the gas is kinematically distributed by the large scale motion 
($>$ 10 pc) 
of HDL.
AFGL\,333 is located on the near side of the W4 bubble, and W3\,Main and W3\,(OH) are at the edge or far side of the bubble,
as suggested by \cite{thr85}.
The inverse trend of the velocity structure near the center may reflect the small scale structure within the clumps. 

\begin{figure}
\begin{center}
\includegraphics[width=8cm]{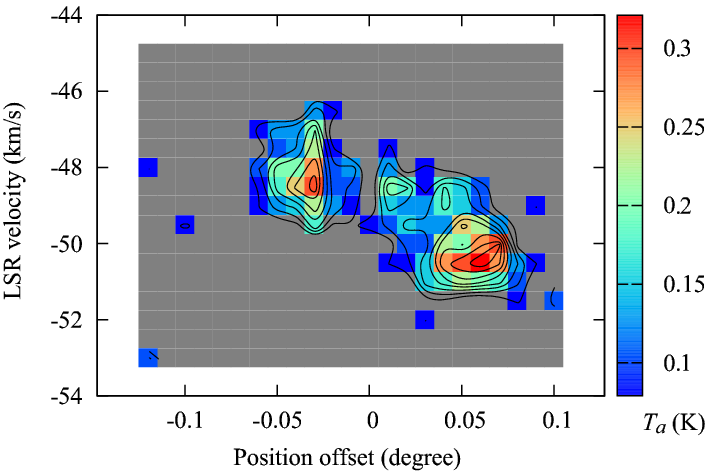}
\end{center}
\caption{Position velocity map of \nh3\ (1,1) line.
         The lowest contour and contour interval are 
         0.08 K (2$\sigma$) and 
         0.03 K, respectively.
         The position offset, increasing toward the northwest, is relative to the position of IRAS\,02245+6115
         along the line shown in figure 4.}
\label{fig:05}
\end{figure}

We found two regions of large velocity width
at the eastern side of the ridge, 
one north and the other south of the dent. 
The (1,1) emission velocity widths of these two regions are 
2.8 $\pm$ 0.4 \kms,
compared to 1.8 $\pm$ 0.1  \kms\ at the other positions.
\cite{jos16} reported YSO surface density peaks at the center of the cluster associated with IRAS\,02245$+$6115.  
Figure \ref{fig:05} shows the position velocity diagram of the (1,1) 
emission line along  the line in figure \ref{fig:04}. 

\begin{figure}
\begin{center}
\includegraphics[width=12cm]{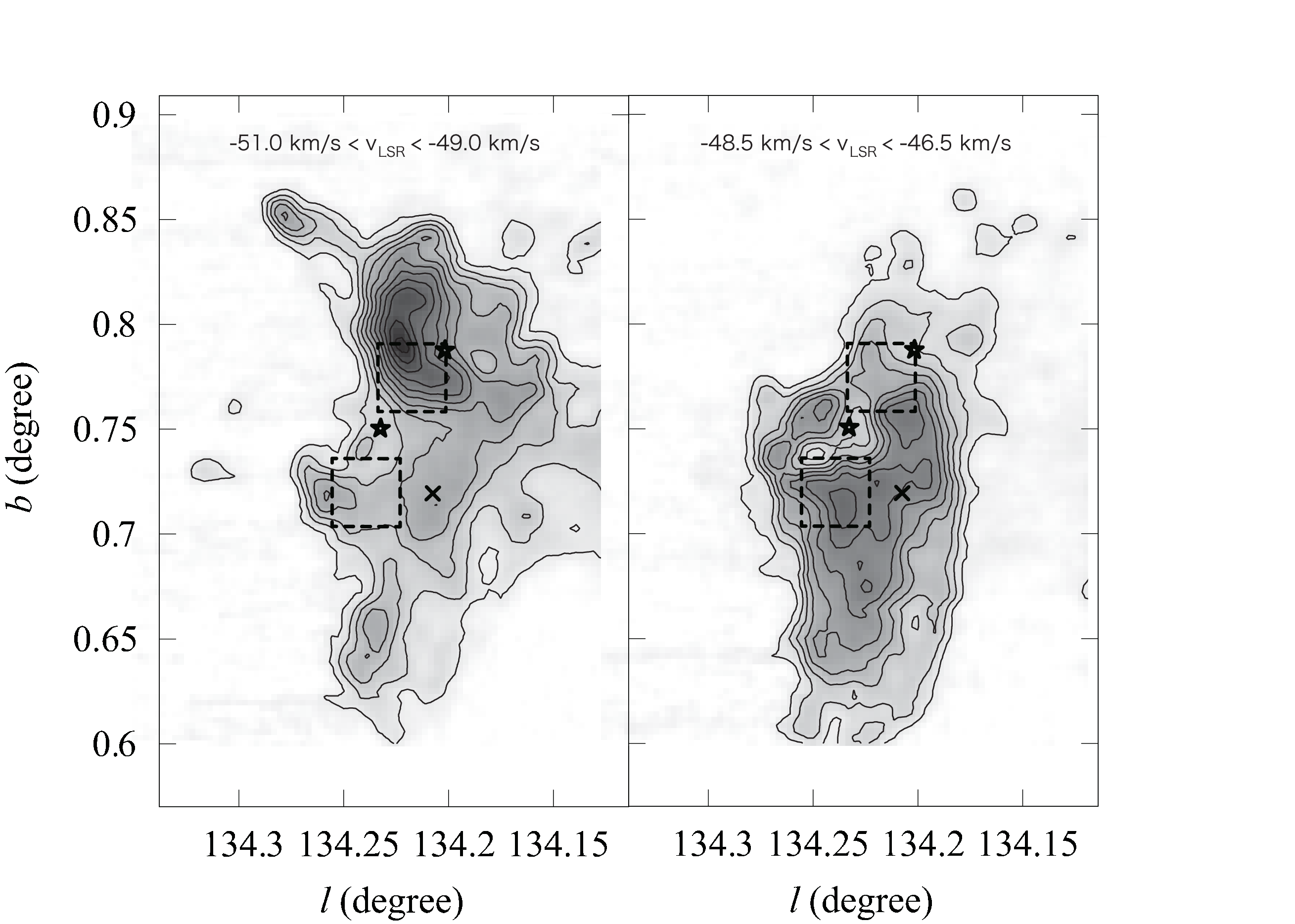}
\end{center}
\caption{$^{13}$CO J=2-1 integrated intensity maps in two velocity intervals, (a) $-51.0 <$ \vlsr $< -49.0$ \kms,
         and (b) $-48.5 <$ \vlsr $< -46.5$ \kms, from the dataset of \cite{bie11}.  Two IRAS sources, large \nh3\ line width regions, and an
         \nh3\ peak position are shown in stars, boxes, and x, respectively.
         The lowest contour and contour interval are 
         2.5 K \kms (5$\sigma$) and 
         2.5 K \kms, respectively.}
\label{fig:06}
\end{figure}

We made the $^{13}$CO J=2-1 integrated intensity maps in two velocity intervals,
which is shown in figure \ref{fig:06}, by using the
data cubes provided by \cite{bie11}.  
The blue shifted components ( $-50.0$ \kms) mainly consist of the curved ridge at the east of IRAS\,02244$+$6117,
and two clumps at the west and the south of IRAS\,02245$+$6115.
The northern \nh3\ large velocity width region corresponds to the part of this curved CO ridge.
The red shifted components ( $-47.5$ \kms) mainly consist of the circular ridge 
with the size of $\timeform{0.05D}$
around IRAS\,02245$+$6115, and
the southern extension.
The southern \nh3\ large velocity width region corresponds to the part of this circular CO ridge.
\afg, we confirmed on the \nh3\ maps, are not so conspicuous on the $^{13}$CO maps.
This may show the difference of the critical density of the molecular species.
\nh3\ emission sampled denser main part of the \afg. 

We discuss the origin of these 
\nh3\ 
large velocity width regions
in subsection 4.2.

\subsection{Physical Parameters}

From the \nh3\ observational results, 
we derived the physical parameters:  optical depth, column density, 
rotational temperature, local thermodynamical equilibrium mass, 
and virial mass of \afg.

\begin{figure}
\begin{center}
\includegraphics[width=8cm]{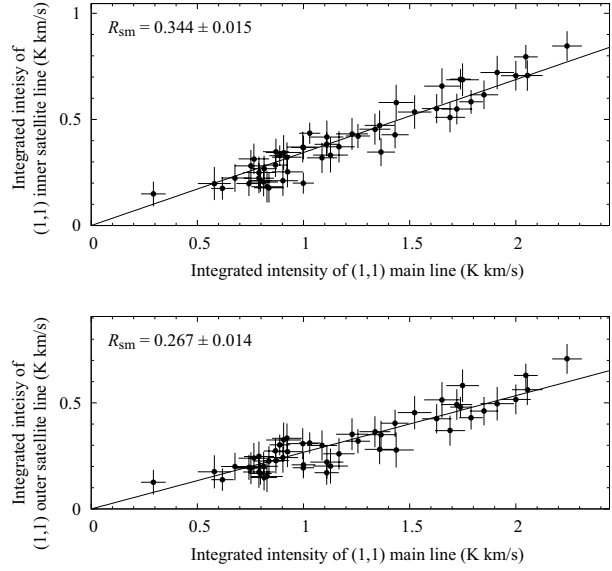}
\end{center}
\caption{Correlations of the integrated intensity in the (1,1) main and
         satellite lines. The correlations of the main to the inner
         and outer satellite lines are shown in the top and bottom panel,
         respectively. Data detected over the $2\sigma$ level in
         both the main and the satellite lines are plotted.
         The error bar shows the $1\sigma$ noise level.
         The estimated intensity ratio is shown in the top left corner
         of each panel.}
\label{fig:07}
\end{figure}

The optical depth of \nh3\ can be derived from the intensity ratio
between the main and satellite line using the method shown in
\cite{ho83}.
The theoretical intensity ratio of satellite to main lines is

\begin{eqnarray}
\frac{T_{\rm a}({\rm main})}{T_{\rm a}({\rm sate})}
= 
\frac{1 - e^{- \tau}}{1 -e^{- a \tau}}
\label{eq:1}
\end{eqnarray}

In the case of the inner satellite line and outer satellite line, 
$a$ is 0.278 and 0.222, respectively.
Figure \ref{fig:07} shows the correlation between the \nh3\ (1,1) 
and the inner/outer satellite lines.
The correlation shows a fairly uniform intensity ratio
for all observed positions.
The ratios of the inner and outer satellite lines
to main line were obtained to be $0.344\pm0.015$ and $0.267\pm0.014$,
respectively.
These correspond to optical depths of 
$0.63\pm0.14$ and $0.50\pm0.15$
which are consistent within the error values.
From the average of these two values, the optical depth
of \nh3\ (1,1) in \afg\ was derived to be $0.57\pm0.10$.


\begin{figure}
\begin{center}
\includegraphics[width=8cm]{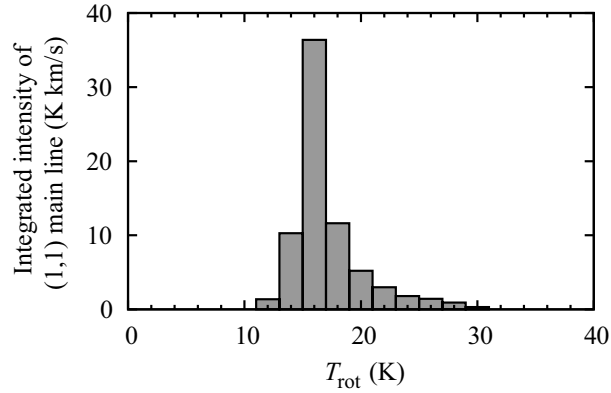}
\end{center}
\caption{Histograms of the rotational  temperature.
         The integrated intensity of the (1,1) main line is each rotational temperature.}
\label{fig:08}
\end{figure}

The rotational temperature is determined from the the intensity ratio of (2,2) 
to (1,1) lines, assuming that the (1,1) and (2,2) lines are emitted from the same gas. 
If the dynamical state and the excitation conditions are similar, the line shapes of 
(1,1) and (2,2) emission should be similar too.  
To check the similality, we made gaussian fitting to the (1,1) and (2,2) lines. 
The obtained central velocity and the velocity width are $-48.67\pm0.03$ \kms\ 
and $1.92\pm0.06$ \kms\ for the (1,1) line, $-48.60\pm0.09$ \kms\ and $1.75\pm0.20$ \kms\ 
for the (2,2) line. They are consistent each other.
In the case that the optical depth is uniform,
the rotational temperature, $T_{\rm rot}$, is directly derived from the
the intensity ratio of 
the (2,2) to (1,1) line.
Following \cite{ho83}, we derived the rotation temperature as:

\begin{eqnarray}
T_{\rm rot}(2,2; 1,1)
= 
-41.5 
\left/
\ln
\left(
\frac{-0.282}{\tau (1,1,m)}
\times
\ln
\left[
1 - \frac{T_{\rm a}(2,2)}{T_{\rm a}(1,1)}
\times
\{ 1 - \exp[-\tau (1,1,m)] \}
\right]
\right)
\right.
\label{eq:2}
\end{eqnarray}

\begin{figure}
\begin{center}
\includegraphics[width=12cm]{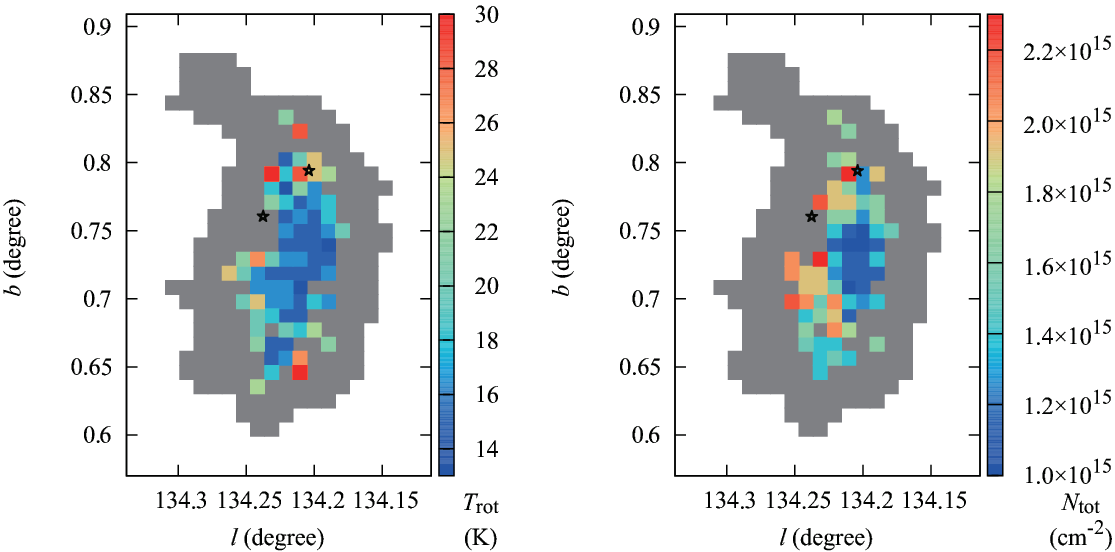}
\end{center}
\caption{Rotational temperature map (left) and column density map (right).
              Stars indicate the positions of two IRAS sources as figure 4.}
\label{fig:09}
\end{figure}

The average intensity ratio is obtained to be $0.336\pm0.020$ from the 
correlation plot 
of the integrated intensity in the (1,1)  main and (2,2) main lines.
However the intensity ratio seems to change in the range from 0.2 to 1.0.
This indicates the rotational temperature is different with 
the observed positions.
Figure \ref{fig:08} shows a histogram of the rotational temperature
for the (1,1) integrated intensity with 
$2\sigma$ detection in both the (1,1) and (2,2) lines.
The (3,3) line was used to discuss the interstellar shock through the abundance of 
the ortho-to-para \nh3. However, the average intensity ratio of (1,1) line to (3,3) line 
in this region is not high ($0.20\pm0.02$) as a typical molecular clouds in the galactic disk \citep{nag09},
implying that the special condition as the interstellar shock is not
necessarily required.
The value of rotational temperature ranges between $12-30$ K.
The mean value and standard deviation of the rotational temperature
were derived to be 17 K and 3 K, respectively.
Scan positions that derived temperatures in the range of $T_{\rm rot} = 12-22$ K and
$23-30$ K account for 94\% and 6\% of the total integrated intensity, respectively.
$T_{\rm rot} = 12-22$ K is consistent with the dust temperature 
estimated from the Herschel data of \afg\ \citep{riv13}.
Figure \ref{fig:09} shows the rotational temperature map.
The gas with a rotational temperature higher than the dust temperature
($23-30$ K) seem to be located at north, east, and south sides of the ridge.
The temperature is enhanced 
nearly 10 K
near the ionization front facing W4.
Such an enhancement is also seen in the excitation temperature of CO
\citep{sak06,pol12}. 

The column density of \nh3\ can be estimated from the column density 
of (1,1) line and the rotational temperature using the assumption of 
Local Thermal Equilibrium (LTE) \citep{man92}.

\begin{eqnarray}
N(1,1) = 2.78 \times 10^{13} \tau(1,1,m) 
\left(
\frac{T_{\rm rot}}{[{\rm K}]}
\right)
\left(
\frac{\Delta v_{1/2}}{[{\rm km\ s}^{-1}]} 
\right)
\rm{\ \ [cm}^{-2}] 
\label{eq:3} 
\end{eqnarray}

The column density of the (1,1) line can be estimated from
the product of the optical depth of the (1,1) line, the rotational temperature, 
and the velocity width (FWHM) of the (1,1) line.
We used a uniform optical depth of $0.57\pm0.10$ for the whole observed area.
We used the rotational temperature and the velocity width obtained at each 
observed positions shown in figure \ref{fig:09} and \ref{fig:04}.

Again, using the LTE assumption, the total column density of \nh3\,
which is sum of the column density of each rotational levels,
was estimated using the column density of (1,1) line and 
the rotational temperature at each observed position.
The column density map is shown in figure \ref{fig:09}.
The mean column density was obtained to be $(1.4\pm0.4)\times10^{15}$ cm$^{-2}$.
The error is the standard deviation of whole observed area.

\begin{figure}
\begin{center}
\includegraphics[width=12cm]{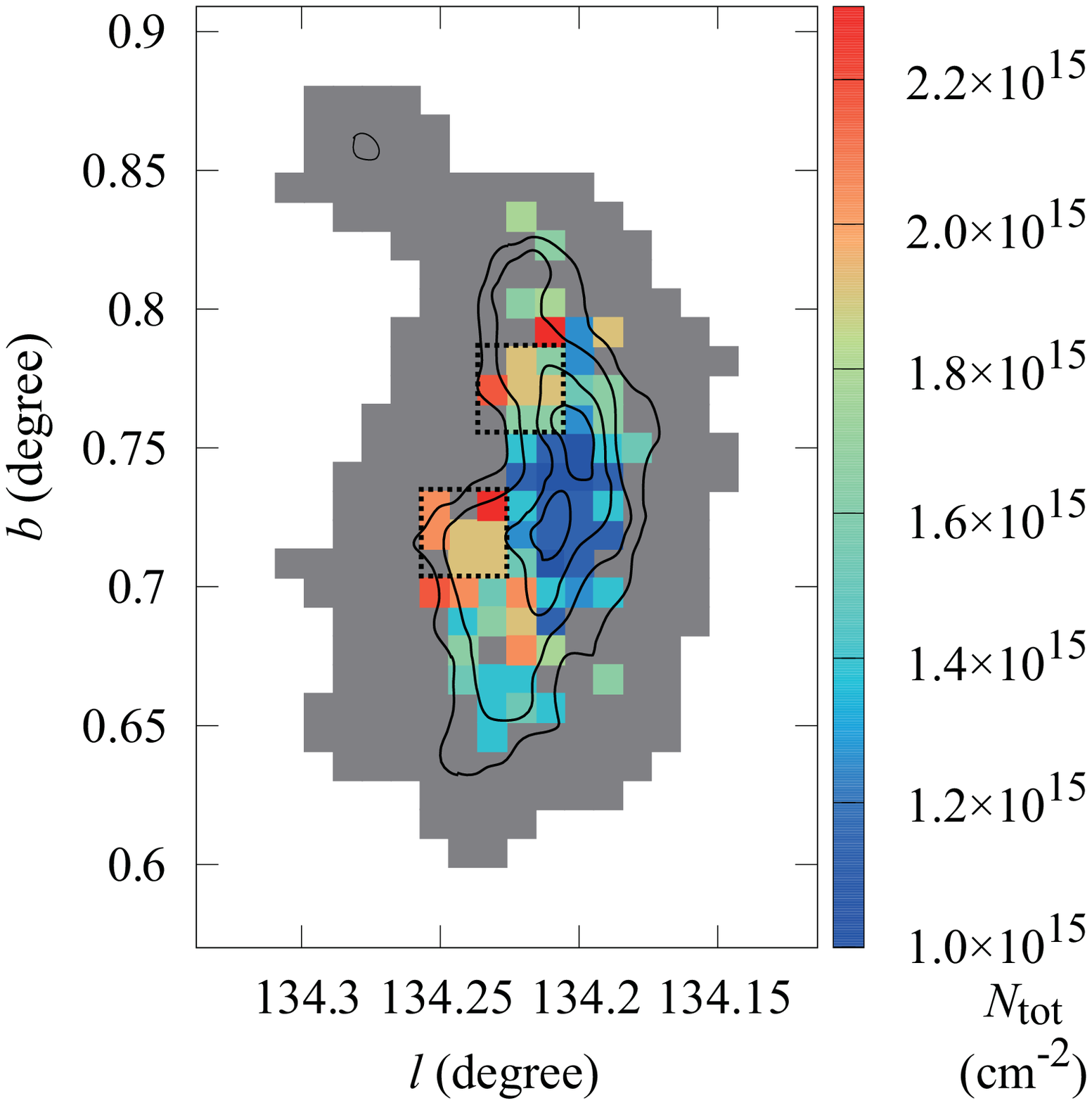}
\end{center}
\caption{H$_2$ column density contour \citep{riv13} overlaid on the \nh3\ column density map.}
\label{fig:10}
\end{figure}

The derived column density is compared with other observational results.  
\cite{mor14} made ammonia observations by a petal-shaped scan centered on sub-mm sources in 
W3 GMC. 
Their six ammonia clumps, W48-W53, are located within our observed region excluding W54,
which corresponds to SFO\,05. 
The column 
densities of their sources,
W48, W49, and W50, 
which are located near the \afg\, is $(5.1\pm0.5)\times10^{14}$ cm$^{-2}$. 
However, for the sources in the eastern side of \afg\ (W51, W52, and W53),  
the column density is $(6.8\pm1.5)\times10^{14}$ cm$^{-2}$.   
Although their values are nearly 
a third
of ours ($10-19\times10^{14}$ cm$^{-2}$), 
the column density is higher at the eastern side of the \afg.
\cite{riv13} showed the H$_2$ column density map of AFGL\,333 region 
by the Herschel HOBYS data. The central column density is $100\times10^{21}$ cm$^{-2}$. 
Figure \ref{fig:10} shows their contours of the H$_2$ column density overlaid on our \nh3\ column
density map.
As the increase of the \nh3\ column density is mainly around the positions of large velocity widths,
one possible cause is the \nh3\ velocity width. 
Another 
cause might be 
due to the actual \nh3\ abundance or excitation conditions within the clumps \citep{mor14}.
If we adopt N(H$_2$) derived from dust continuum \citep{riv13}, 
the abundance ratio of \nh3\ relative to H$_2$, $X({\rm NH}_3)$,  is $3\times10^{-8}$ at the eastern edge 
and  $1\times10^{-8}$ at the central ridge.

The LTE mass of the cloud, $M_{\rm LTE}$, 
was estimated from the sum of the LTE mass obtained at each observed grid, 
$M_{\rm g}$, using 
$M_{\rm LTE} = \sum M_{\rm g} = \sum S_{\rm g} N({\rm NH}_3)_{\rm g} m({\rm H}_2) / X({\rm NH}_3)$,
where $S_{\rm g}$ is the grid area of $37.5\arcsec\times37.5\arcsec= 0.13$ pc$^2$,
$N({\rm NH}_3)_{\rm g}$ is the column density of \nh3\ 
at each observed grid shown in figure \ref{fig:09},
$m({\rm H}_2)$ is the mass of the hydrogen molecule.
The fractional abundance of \nh3\ has typical values of $10^{-9}$ to $10^{-7}$ in
dense gas clumps \citep[e.g.][]{ung80, ho83, tif98}.
By assuming $X({\rm NH}_3)$ of 3$\times10^{-8}$ as in the case of
other massive star forming regions \citep[e.g.][]{dun10,urq15}  and
infrared dark clouds \citep{chr13}, 
the LTE mass of the cloud is $7700\pm1000 M_{\odot}$.

The other cloud mass estimation is the virial mass, $M_{\rm vir}$.
This is derived from $M_{\rm vir} = 250 R \Delta v^2$,
where R is the radius of the cloud and $\Delta v$ is the
velocity width \citep{roh96}.
The virial mass within the radius of 1.6 pc ($\timeform{0.10D}$),
which corresponds to the same area of the LTE mass estimation,
was estimated to be $1800\pm400 M_{\odot}$.
In this estimation, the mean velocity width of whole observed area
of $2.15\pm0.46$ \kms\ was used. 
Our results suggest that the \afg\ is not close to a virialized state.
\cite{sak06} identified two cores, A and B, in the \afg\
from their C$^{18}$O observations.  They derived the LTE masses and the viral masses of two cores, 2500 and 900 $M_{\odot}$,
and 1400 and 500 $M_{\odot}$,  for core A and B, respectively.

\subsection{New \h2o\ Maser Detection}

\begin{figure}
\begin{center}
\includegraphics[width=8cm]{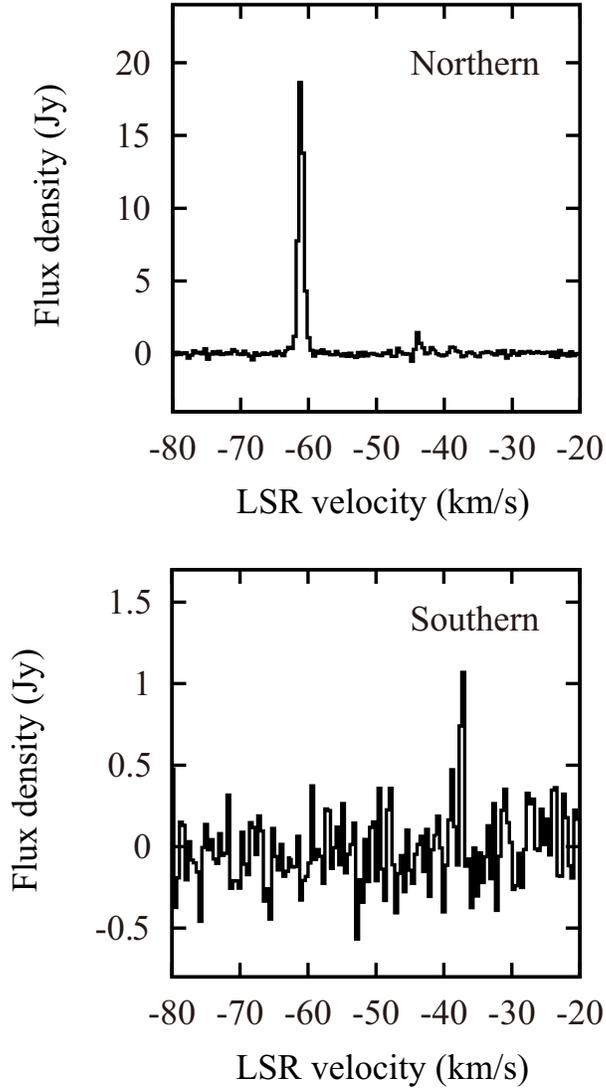}
\end{center}
\caption{\h2o\ maser spectra of the northern and the southern masers. 
         Detected positions are listed in Table \ref{tab:01}.}
\label{fig:11}
\end{figure}

In star forming regions, \h2o\ masers serve as signposts of sites of 
active star formation. 
They may trace shocks, outflows, and other episodic events 
during the early 
formation
stages of a protostar 
\citep{chi12, chi14, tor11}. 
Observing \h2o\ maser emission simultaneously to \nh3\
lead us to detect \h2o\ masers at two different 
location in the region. 
For identification purposes, we will refer to these masers as 
the northern-maser and the southern-maser. 
The northern-maser is associated with SFO\,05 and 
is a known maser which was reported by 
the maser survey of bright rimmed molecular clouds
\citep{val08}. 
However, the position around the southern-maser was not searched before and not listed in any maser catalogs 
\citep[e.g.][]{val05, sun07}.
The southern-maser is located near the center of \afg.
Figure \ref{fig:11} shows the two \h2o\ maser spectra.
The positions, radial velocities, and peak fluxes with uncertainty of the masers are shown in Table \ref{tab:01}. 
The error in their determined positions is 30\arcsec.


\begin{table}
  \tbl{\h2o\ Masers detected in our mapping observation.}{%
  \begin{tabular}{lrrrrrr}
      \hline\hline
      Maser & $l$ & $b$ & $\alpha_{\rm 2000}$   & $\delta_{\rm 2000}$   & \vlsr (\kms)  &  Peak flux (Jy)  \\
      \hline
      Northern    & $\timeform{134.2790D}$ & $\timeform{0.8571D}$ & $\timeform{02h29m02.0s}$ & $ \timeform{+61D33'35''}$ & $-61\pm0.4$ & $18.44\pm0.12$ \\
      Southern   & $\timeform{134.2069D}$ & $\timeform{0.7291D}$ & $\timeform{02h28m04.7s}$ & $ \timeform{+61D28'01''}$ & $-36.9\pm0.4$ &  $1.06\pm0.14$ \\
      \hline
    \end{tabular}}\label{tab:01}
\end{table}

\begin{figure}
\begin{center}
\includegraphics[width=12cm]{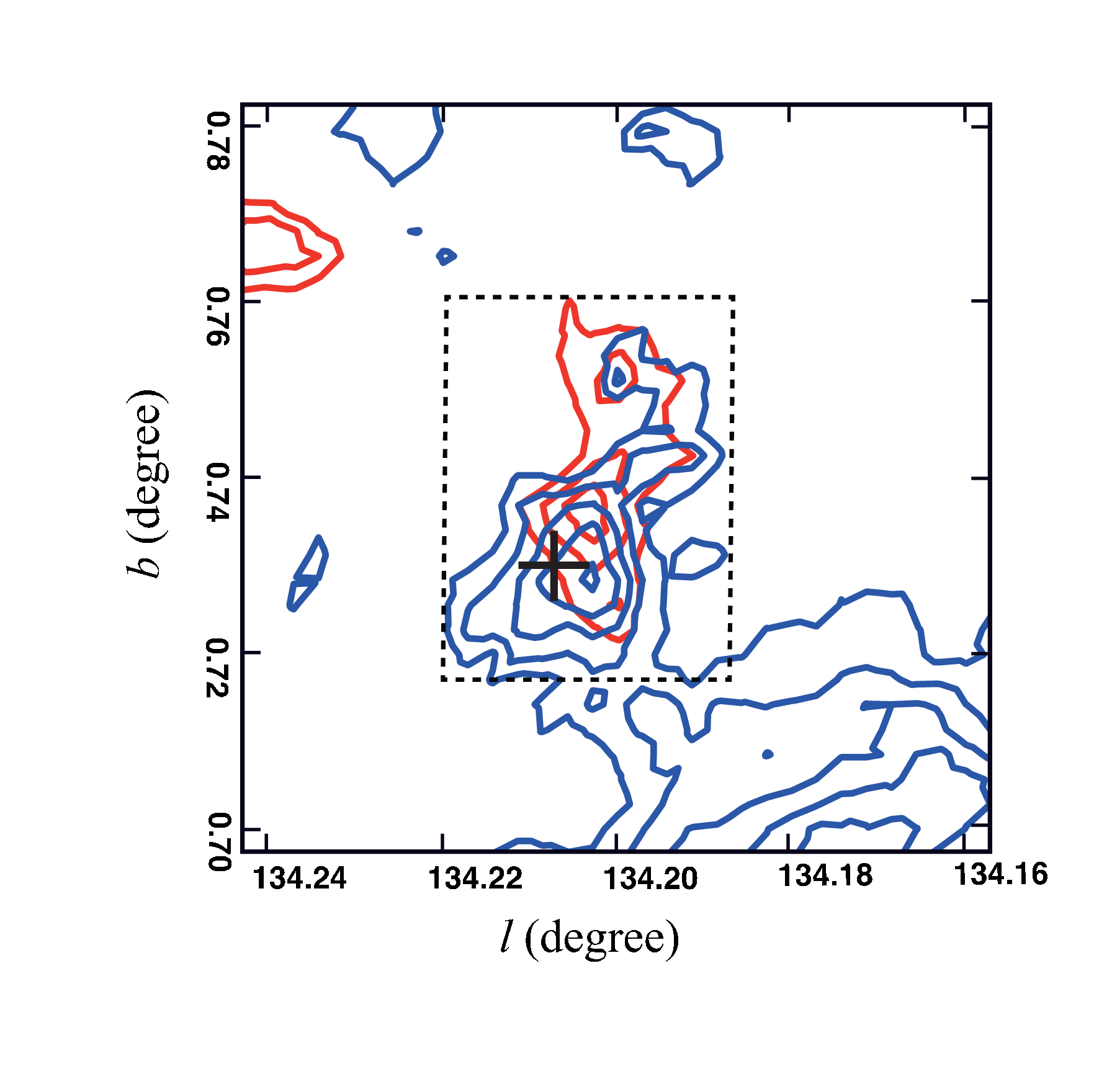}
\end{center}
\caption{Contour plot of $^{12}$CO J=2-1 red and blue outflow lobes.
     Velocities between $-62.0$ and $-54.0$ \kms\ are shown as blue contours and velocities
     between $-43.0$ and $-35.0$ \kms\ are shown in red contours.
     The blue component at the southwestern part is probably the contamination
     unrelated to the outflow associated with the young stellar activity.
     The lowest contour and contour interval are $3.4$ K \kms\ (2$\sigma$) and $3.4$ K \kms.
     A plus marks the location of the southern-maser.
     The dashed box shows the region where the physical parameters of the outflow are calculated. 
 }
\label{fig:12}
\end{figure}

\begin{table}
  \tbl{CO outflow parameters associated the southern \h2o\ maser.}{%
  \begin{tabular}{lccc}
   \hline\hline
   Lobe & M [$M_{\odot}$] & P [$M_{\odot}$ km s$^{-1}$] & E [$M_{\odot}$ km$^{2}$ s$^{-2}]$\\  
   \hline
   Blue   & $0.76$ & $4.97$ & $32.4$ \\
   Red   & $0.49$ & $3.02$ & $17.3$ \\
   \hline
  \end{tabular}}\label{tab:02}
\end{table}

We found a high velocity  $^{12}$CO wing components, which suggest the activity of young stellar object.
Figure \ref{fig:12} shows the distribution of the emission wings integrated over the velocity  range 
$-45$ \kms\  $<$ \vlsr\ $<  -37$ \kms\ (red component), and 
$-60$ \kms\  $<$ \vlsr\ $<  -52.5$ \kms\ (blue component) around
the southern-maser.  The blue components at the southwestern part of the map is probably
another velocity ($-53$ \kms)  component. Our result suggests the bipolar molecular outflow
is associated with the southern-maser. 
Assuming $^{12}$CO J=2-1 line is optically thin, and CO abundance relative to H$_2$ 
is 10$^{-4}$, the lower limit of the outflow masses in the blue and red lobes.
We restricted the area of the outflow within the box area of 150\arcsec\ 
in latitude and 120\arcsec\ in longitude, shown in the dashed box in figure \ref{fig:12}.
As the temperature of the warm gas in the outflow is uncertain,
we choose T$_{ex}=17.6$ K to minimize the column density as  \cite{dun10b}.
The calculated total mass,  momentum and energy of the outflow in each lobe are
shown in Table \ref{tab:02}. 
There are four Class I  
and one Class II sources
\citep{jos16} within the positional accuracy  
of the southern-maser 
(table \ref{tab:03}). 
All five sources were detected by the Spitzer IRAC bands, but have no NIR data, and
two sources  were detected by MIPS 24 ${\mu}m$ band.
It suggests that the outflow is powered by the source(s) among them.

\red{
\begin{table}
  \tbl{
  YSO candidates near the southern \h2o\ maser, taken from Table 3 of \cite{jos16}.
  }{%
  \begin{tabular}{lllllrrrrcc}
   \hline\hline
   ID & $\alpha_{\rm 2000}$ & $\delta_{\rm 2000}$ & $l$  & $b$ & [3.6] & [4.5] & [5.8] & [8.0] & [24] &YSO Classification\\ 
        & (deg)                    & (deg)              & (deg) & (deg) & (mag) & (mag) & (mag) & (mag) & (mag)& \\
   \hline
   1 & $37.0250$ & $+61.4701$ & $134.2082$   & $0.7329$ & 12.83 & 10.42 & 8.97 & 7.89 & 3.04 & Class I\\
   2 & $37.0258$ & $+61.4689$ & $134.2090$   & $0.7319$ & 12.83 & 11.18 & 10.02 & 9.20 & ... & Class I\\
   3 & $37.0227$ & $+61.4686$ & $134.2077$   & $0.7311$ & 12.26 & 10.61 & 9.39 &8.54 & 2.59 & Class I\\
   4 & $37.0159$ & $+61.4634$ & $134.2066$   & $0.7251$ & 14.96 & 13.96 & 12.82 & 11.98 & ...  & Class I\\
   5 & $37.0243$ & $+61.4644$ & $134.2100$   & $0.7275$ & 14.46 & 13.40 & 13.15 & 11.37 & ... & Class II\\
   \hline
  \end{tabular}}\label{tab:03}
\end{table}
}

\section{Discussion}
\subsection{Star Formation Activity in the Region}

There have been extensive studies on W3 GMC, but mainly on the most active star forming regions such as W3\,Main and W3\,(OH).
For example,  the deep NIR imaging of the W3\,Main, \cite{bik14} revealed the disk fraction of the young stellar content. 
The age of IC\,1795 OB association was estimated to be 3-5 Myr  by \cite{oey05} and \cite{roc11}.
\cite{rom15} identified five principal clusters and discussed the history of star formation in W3\,Main and W3\,(OH). 
Across the W3 region, extensive survey of the high-mass stellar population by \cite{kim15}  showed star formation in W3 began spontaneously up to 8--10 Myr ago.
\cite{riv15} made dust column density maps by using Herschel datasets in the whole W3 complex, and created the probability density functions (PDFs). 
They analyzed PDF to show the stellar feedback-based constructive process, i.e. stellar feedback is a major player in the cluster formation 
and the overall characteristics and local evolution of GMC. 
They proposed a ``convergent constructive feedback" model.
They suggested the combined collection effect in compressing and confining material
by the high mass stars create higher column densities by feedback, which lead
to the creation of an environment suitable for high mass cluster formation.

\begin{figure}
\begin{center}
\includegraphics[width=17cm]{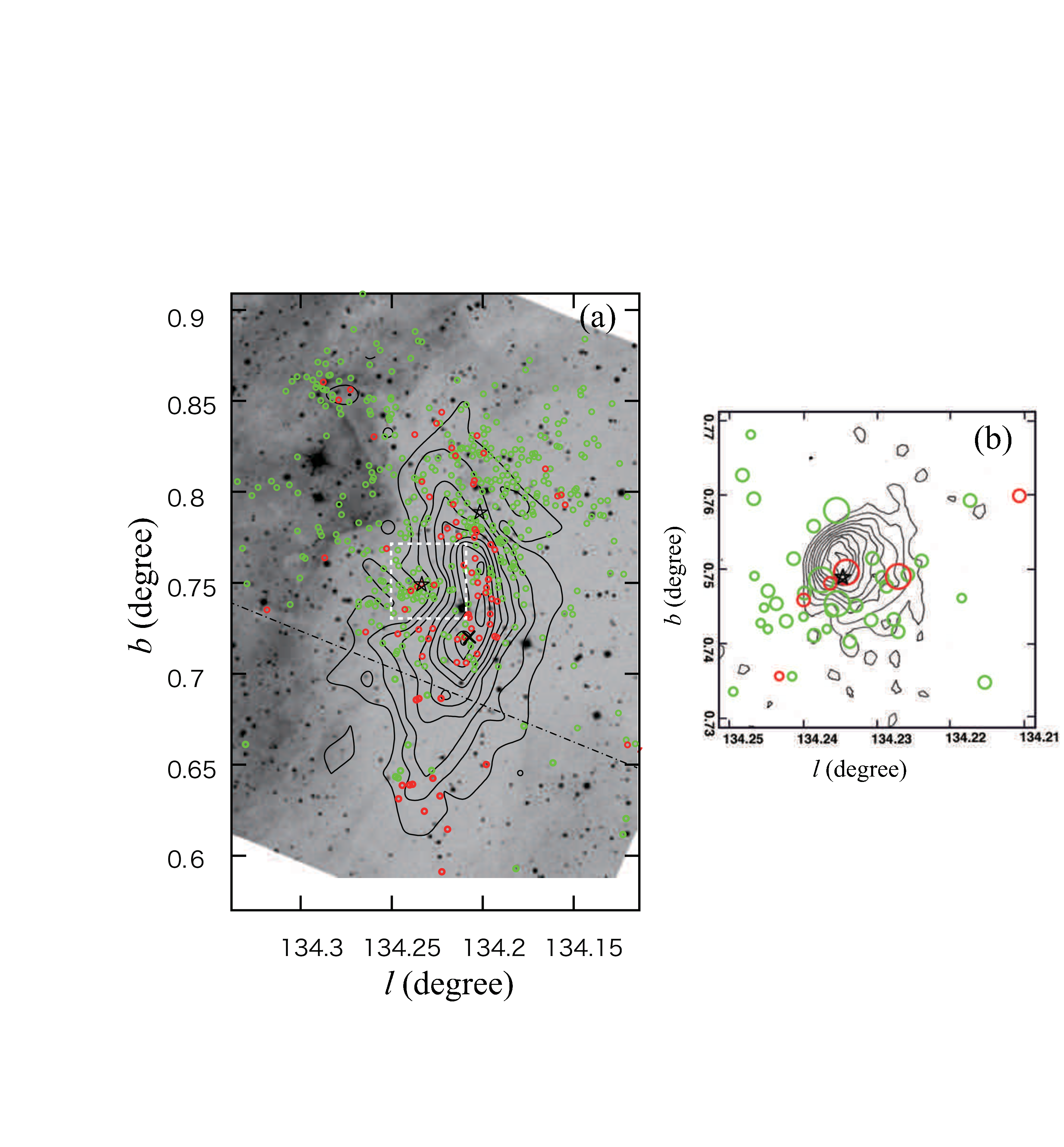}
\end{center}
\caption{(a) Distribution of YSOs and the contour map of \nh3\ integrated intensity overlaid on the DSS-R image.  
       Class I, Class II sources \citep{jos16} are shown in red and green circles, respectively.
       In the southern part below  the dash-dotted line, Class 0/I and Class II sources
       by \cite{riv11} are shown.  Two IRAS sources and an \nh3\ peak position are shown in stars and x, respectively.
       (b) YSOs with NIR color data in the box of $\timeform{0.04D}$ around IRAS\,02245$+$6115, which is shown with white dashed line in (a),
       are overlaid on the synthesis radio continuum map \citep{hug82}. 
       YSOs are shown by the symbols in different size according to their mass estimated based on the color magnitude
       ( $J-H$ vs $J$ ) diagram by \cite{jos16}.  
       Sources with  $M < 1.0 M_{\odot}$, $M = 1.0 - 10.0 M_{\odot}$, and $M > 10 M_{\odot}$ are shown 
       in small, medium, and large circles, respectively.          
       }
 \label{fig:13}
\end{figure}

Recently, \cite{jos16} made deep JHKs photometry 
of the AFGL\,333 region, complementing
with Spitzer IRAC and MIPS
observations. Their new NIR photometry is $> 3$ mag deeper in each band compared to the  2MASS photometry,
and they identified many YSOs including low mass populations in this region. 
To explore the distribution of YSOs in our observational field, we used 
their candidate sources of Class I and II.
Figure \ref{fig:13}a shows the plot of distribution of the YSOs.
As their field is not fully covered our mapping area of \nh3\ observations,
we supplemented by the Spitzer YSO catalog (Catalog 1) of \cite{riv11}.
That is why the surface density of Class II sources in the southern region appears to be low. 
A rich clustering of Class 0/I objects is seen along \afg\ and in the core region, where
the southern-maser source and \nh3\ peak lie.
Although in the Spitzer Catalog 1 clusterings are not found around the positions of two IRAS sources
because of confusion by the bright infrared nebulosities, 
the deep NIR observations by \cite{jos16} clearly revealed some peaks of the YSO surface density 
associated with \afg\ and its surroundings. 
The peak around IRAS\,02245$+$6115 was 
already discussed by \cite{car00}.  \cite{bic03} also listed it with the size of
1\arcmin.7 $\times$ 1\arcmin.5 in the catalog of infrared clusters and stellar groups.
IRAS\,02245+6115 is associated with 
compact \hii\ region G\,134.2+0.8 \citep{hug82} 
which has an electron density of 450 cm$^{-3}$ and size of 0.3 pc.
The  peak of the 5 GHz continuum emission agrees well with IRAS\,02245$+$6115
within a positional uncertainty of $27{\arcsec} \times 9{\arcsec}$ in  position angle $58^{\circ}$
(see figure \ref{fig:13}b).
One of the Class I source at $(l, b) = (\timeform{134.2344D}, \timeform{0.7497})$, which also corresponds to
this peak, have $J = 15.98$,  $H = 13.34$, and $K_s = 11.71$ \citep{jos16}.
The extinction map made from the CO column density \citep{jos16} indicates $A_v \sim$ 25 mag around G\,134.2+0.8.
On the NIR color-magnitude diagram (figure 8 by \cite{jos16}), 
its location is consistent with deeply embedded star with the mass  of more than $10 M_{\odot}$ using 
the 2 Myr isochrone by \cite{bre12}.
Thus, this source should be the exciting star of the compact \hii\ region as suggested by \cite{mam84}.
Most YSOs are distributed in the south-western of this peak of radio continuum emission.
Another IRAS source, IRAS\,02244$+$6117, is a bright infrared source  BIRS-104 \citep{elm80} and
a small arc of infrared emission around it is reported by \cite{kra03}.
The peak of YSO surface distribution at the south of IRAS\,02244$+$6117 is also evident.
As the northern part of \afg\ is located between these two IRAS sources, \cite{riv11} suggested 
the star formation activity of \afg\ has been induced by these two sources.
The  substructure of \afg\  could be 
affected 
by the feedback from the young stellar activities. 

SFO\,05  at the northern end of \afg\ is the 
typical case of star formation triggered by radiation-driven implosion.
A pre-existing condensation of about $400 M_{\odot}$ was compressed 
by the pressure of the ionizing radiation of W4 \citep{fuk13}.
And the compact \hii\ region G\,134.2+0.8 lies at the interface between the 
\afg\ and W4.
This implies that the B0.5 star exciting the \hii\ region G\,134.2+0.8 
\citep{hug82}
could have been formed due to the influence of the expanding bubble of W4. 
The large-scale temperature enhancement in the eastern side of the \afg\ suggests
an interaction between the dense ridge and the W4 bubble.
The interaction with the compact \hii\ region may be limited to only 
the periphery or the north-south direction where large velocity 
widths were observed.
The low ionizing flux can erode the cloud, but would not trigger the formation of stars \citep{bis11}.
Previous studies \cite[e.g.][]{oey05, lef97} support a triggered 
or induced 
mode of star formation to be responsible 
for the formation of the stars at the interface between the W4 and 
the dense cloud, \afg. 
We suggest the overall structure of \afg\ has been made by the external feedback
from W4.  Although the local stellar feedback was also acting to make small-scale structure of the cloud, 
the convergent constructive process \citep{riv13} did not work effectively in this region. 

Quiescent 
mode of
star formation can be described as a mode of 
star formation with minimal external influence. 
Based on the evidence seen in small velocity widths 
in the \nh3\ lines around this region, the younger generation of YSOs distributed around the center of 
the peak of the \nh3\ (1,1) emission at the core of \afg\ 
may have formed spontaneously.
\cite{dal15} gave a warning in distinguishing triggered population from spontaneously formed ones,
and interpreting the observational data of star forming region in terms of triggering.
Although our observations show  G\,134.2+0.8 
and the W4 bubble are interacting with \afg\ and has affected its 
structure, it is not evident that most stars in this region were formed by the 
induced
mode.
\cite{jos16} compared the star formation efficiency and rate with W3 Main and the nearby low-mass 
star forming regions, and showed the star forming activity in AFGL\,333 region is comparable to other low-mass regions.
As the mass of the cloud estimated by them is consistent with ours, 
the star formation efficiency (SFE $= M_{star} / (M_{star} + M_{cloud}$)) is as low as 3\%.
The stellar feedback has not globally 
enhanced
the star forming activity in this region.
Our finding of the small velocity width of \nh3\ line
in the \afg\  and the existing literatures as an older population in the western side 
of the AFGL\,333 region 
suggest apparent coexistence of induced and quiescent modes of star formation within several parsec scale. 


\subsection{Interaction Between compact \hii\ Region G\,134.2+0.8 and Dense Molecular Cloud}

\nh3\ clumps associated with active star forming regions exhibit large emission line velocity widths 
driven by inflow or outflow within the gas, other non-thermal turbulent motion, 
or numerous unresolved dense clumps \citep[e.g.][]{hin10, urq11}.
In the case of Gem OB1 cloud, the dense cloud in contact with 
an expanding \hii\ region showed no obvious 
observable interaction between them \citep{chi13}. 
On the other hand, the observed large velocity widths 
found in the northern and 
southern surroundings of \hii\ region G\,134.2+0.8
(see figure \ref{fig:04}) is evident, and these feature suggests 
G\,134.2+0.8  is interacting with the dense molecular gas of \afg.
No \nh3\ emission was detected in and to the east of G\,134.2+0.8, 
but the \nh3\ emission detected to the west of it 
(around the center of \afg) showed no significant velocity width. 
The surface of the cloud might be irradiated by UV photons from the G\,134.2+0.8. 
Based on the Lyman continuum photon number from  B0.5 exciting star of 
\hii\ region G\,134.2+0.8 at 1 pc from \afg\ is 8 $\times$ 10$^{8}$ cm$^{-2}$s$^{-1}$.
Even 
at the distance of 16--20 pc from seven 
exciting  stars of W4 \citep{lef97}, the ionizing flux   
is 4 $\times$ 10$^{9}$ cm$^{-2}$s$^{-1}$ at \afg. 
These photon flux is approximately one order of magnitude smaller than for 
the typical star forming clouds at the edge of  \hii\ bubbles  described in the
collect-and-collapse model \citep[e.g.][]{deh03}.
The activity of young stars embedded in the cloud might be a source of 
turbulent motion, but there is not so many young objects in these regions.
One explanation for the north-south alignment 
of this interaction could be an injection of mechanical energy from YSO jets
or high-velocity outflows located near the center of
G\,134.2+0.8, interacting with the dense cloud around it. 
Although, such shock activities provide possible explanation for the observed 
large velocity widths, CO observations by \cite{sak06} show no sign
of a high velocity component expected in the case of such outflows.
The systematic outflow survey of IRAS sources by \cite{sne90} also reported
negative results for the AFGL\,333 region.
However, we should notice that CO observations by \cite{sak06} were not so 
deep to detect faint high velocity wing, because their work was to study the physical 
and chemical states of the cloud.  And the central mapping 
position by \cite{sne90} was different from IRAS\,02245$+$6115.

\begin{figure}
\begin{center}
\includegraphics[width=12cm]{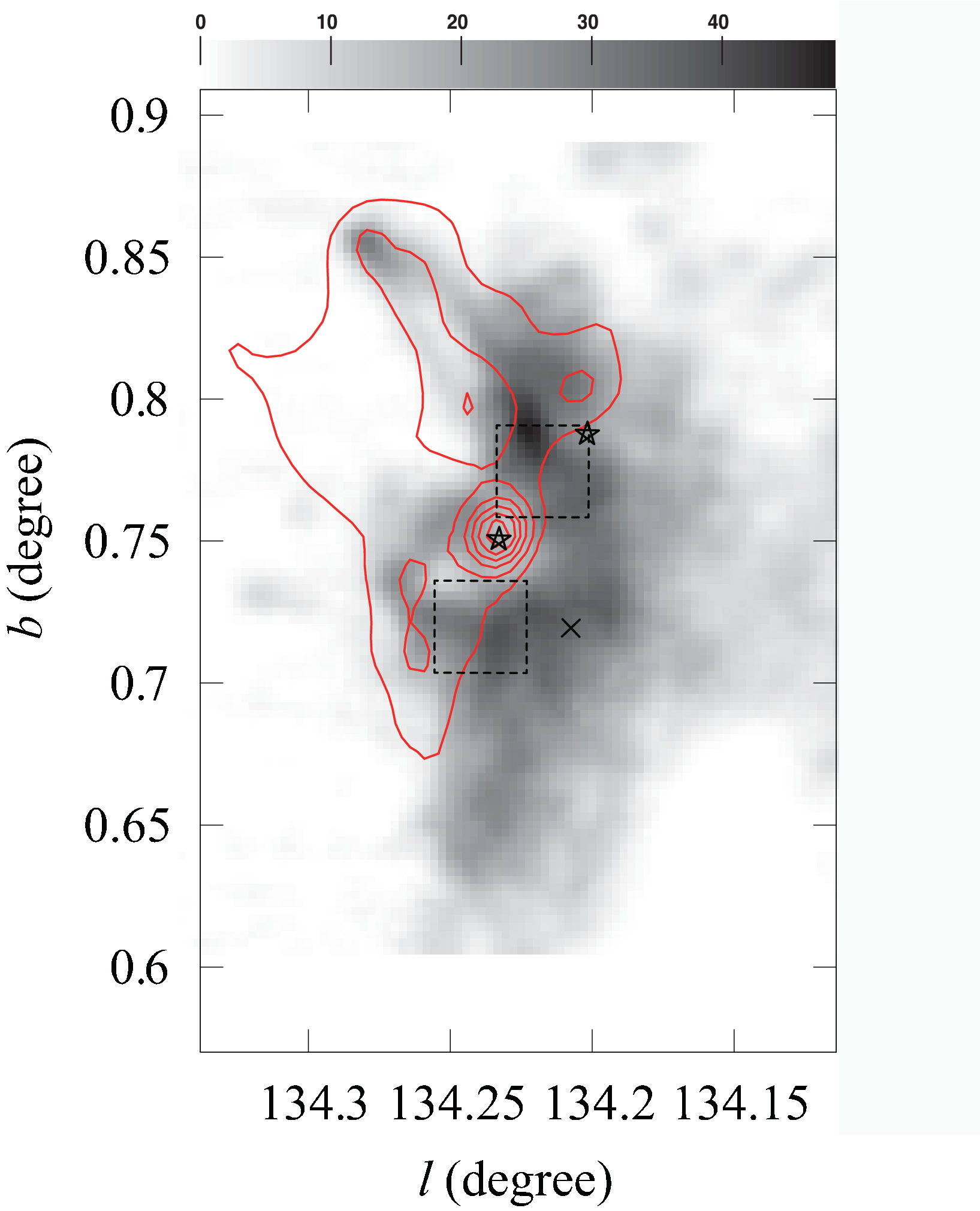}
\end{center}
\caption{The regions of large  \nh3\ velocity width are shown in the dashed boxes
   on the red contours of the 1420 MHz radio continuum emission 
   (Canadian Galactic Plane Survey; \cite{tay03}), with contours incremented 
   in 2 K intervals beginnings with 18 K in brightness temperature, and the gray-scale map
   of integrated $^{13}$CO J=2-1 emission. 
   Two IRAS sources and an
    \nh3\ peak position are shown in stars and x, respectively.}
\label{fig:14}
\end{figure}

Figure \ref{fig:14} shows two large velocity width regions of \nh3\ emission on the
contours of 1420 MHz radio continuum \citep{tay03} and the integrated $^{13}$CO 
J=2-1 intensity map. 
The radio continuum emission of G\,134.2+0.8 fits into the hole of the  $^{13}$CO map.
The cavity or shell-like structure of  $^{13}$CO emission 
with the size of 2\arcmin\ (1.2 pc)  hosting the dense ionized gas 
is reminiscent of the molecular dissociation or dynamical effect of the  \hii\ region on the molecular cloud.
There is the submillimeter source ID-298 \citep{moo07} at the eastern side
of the compact HII region, which corresponds to the CO clump. 
The synthesized radio continuum map  of G\,134.2+0.8 \citep{hug82} shows steep decline of intensity
toward the eastern direction (see figure \ref{fig:13}b).
This asymmetric structure of radio continuum map suggests 
that  a B0.5 star has been formed near this CO clump, and the compact HII region is expanding to the western direction.

\begin{figure}
\begin{center}
\includegraphics[width=15cm]{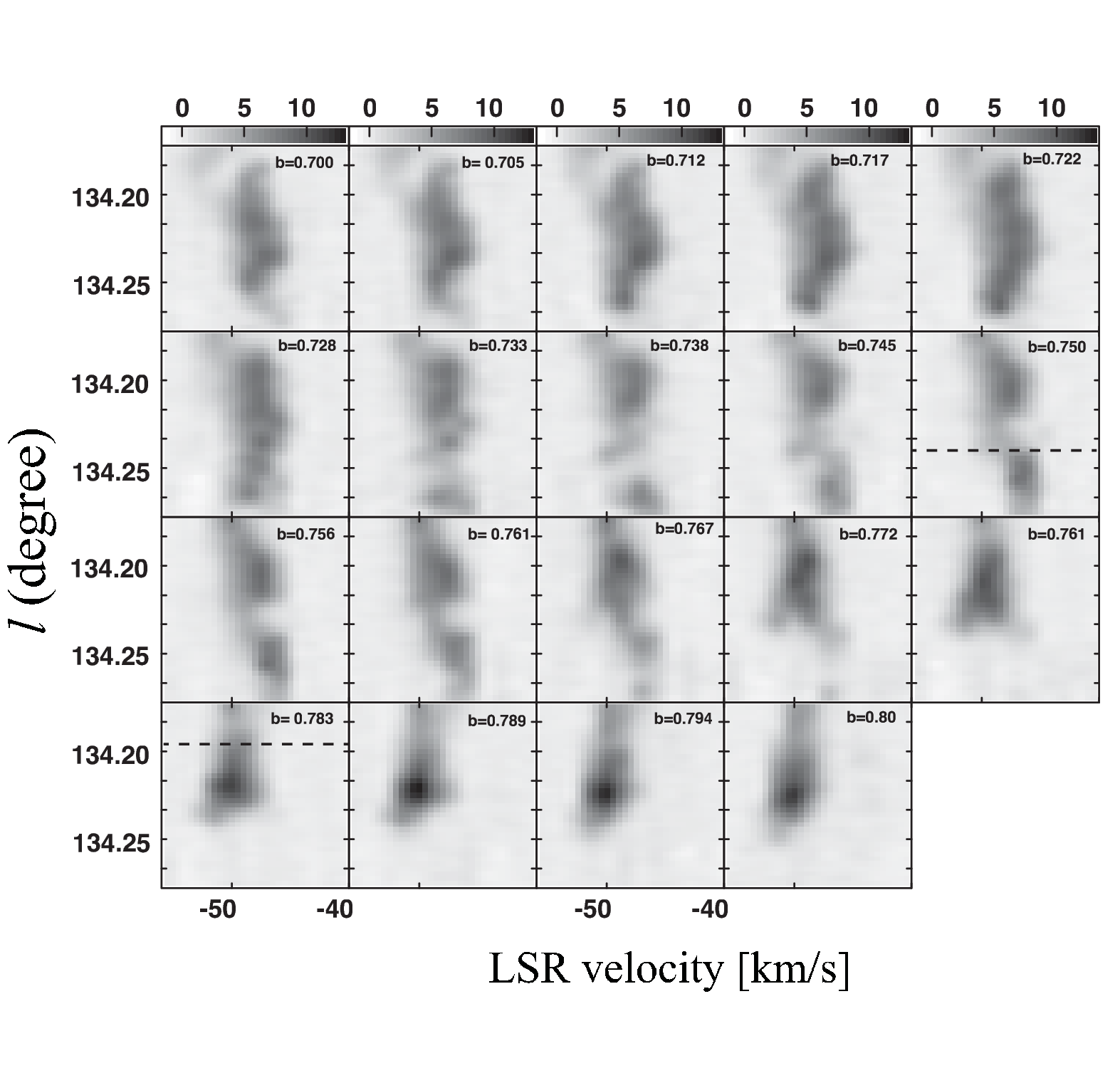}
\end{center}
\caption{Position-velocity diagrams of $^{13}$CO J=2-1 line.
    The positions of IRAS\,02245$+$6115 and  IRAS\,02244$+$6117 are shown in 
    the dashed lines.}
\label{fig:15}
\end{figure}

Figure \ref{fig:15} shows the position-velocity diagrams of $^{13}$CO J=2-1 emission along the galactic longitude.
The positions of IRAS\,02245$+$6115 and IRAS\,02244$+$6117 are shown by the dashed lines 
in the panel of $b = \timeform{0.750D}$ and $b = \timeform{0.783}$, respectively. From $b = \timeform{0.750D}$ to $b = \timeform{0.761D}$, we can trace the C-shaped
velocity structure (blue shifted to $\sim -49$ \kms) centered around $l = \timeform{134.23D}$.  
Such a structure  also shows the evidence of the shell with the expanding motion of $\sim 2$ \kms.
Contrary to the north and south of  G\,134.2+0.8,
the reason for the \nh3\ velocity width of $1.8$ \kms at the west, nearly as low as the main part of the cloud, 
might be the geometrical effect. 
The most dense part of the \nh3\ ridge is deeply embedded in \afg\ along the line of sight.
G\,134.2+0.8 is mostly expanding into the west  at the foreground of the \nh3\ ridge, and could not interact in the western boundary
of the compact \hii\ region.
The interaction is limited in the northern and the southern less dense part of the \nh3\ ridge.
The large velocity width around $( l, b) = (\timeform{134.24D}, \timeform{0.761D})$ in figure \ref{fig:15} 
corresponds to the northern large width \nh3\ region.

Although SED  of the exciting star candidate (see subsection 4.2)
suggests that it is Class I source \citep{jos16}, 
there is no other characteristic signatures of  young activity.
Furthermore, because of its relatively low density, G\,134.2+0.8 is not as young  \citep{thr79}
as the other compact or ultracompact \hii\ regions, like G\,133.8+1.4 (W3\,N) and  W3\,(OH)  \citep{car00},
associated with infrared sources in the W3/W4 region.
For the large line width regions of \nh3\ emission, the explanation invoking interaction by the expanding 
motion of the ionized gas at the periphery of \afg\ is more preferable at present. 
The large line width  could be due to 
turbulent motion supplied through the  shock front of G\,134.2+0.8 in the 
low ambient density. 
Molecular line observations with higher resolution is needed to clarify the activity of the YSOs
in G\,134.2+0.8.

\section{Conclusions and Summary}

Through our \nh3\ mapping observations of  AFGL\,333 in the W4 region
using the Nobeyama 45m radio telescope and archival CO and other data, 
we have explored the physical properties of this region
and studied YSOs in two IRAS sources.
Our results and conclusions are summarized as follows:

1.  We derived a size of 
$2.0 \times 0.6$  
pc, the rotational temperature of $12-30$ K, and 
LTE mass of $7700\pm1000 M_{\odot}$ for the \afg\ molecular cloud.
The viral mass was estimated to be $1800\pm400 M_{\odot}$.  
Possibly indicating that the \afg\ is not close to a virialized state.
We found two regions of large 
\nh3\ line
width at the eastern side of \afg.

2.  The large \nh3\ line width at the north and the south of 
IRAS\,02245$+$6115 suggests an
interaction by the expanding motion of the ionized gas 
from the compact HII region G\,134.2+0.8 at the periphery of 
\afg. $^{13}$CO data obtained by  HHT and NRO reveal a cavity
structure with the size of 1.2 pc in the integrated intensity maps, with an expanding 
motion of a few \kms\ in the position velocity maps.
However, the small velocity widths in \nh3\ gas observed west of
IRAS\,02245$+$6115, around the center of the dense molecular gas, 
suggest that interaction with the compact \hii\ region is limited.

3. Two \h2o\ masers were detected.
The southern-maser at the core of \afg\ was a new finding. 
A CO bipolar outflow, a signpost of active star formation, was found near this maser source
and it appears to be associated with some 
Class I/II
sources.

4. We examined the distribution of YSOs in the AFGL\,333 region, especially around 
IRAS\,02245$+$6115 and IRAS\,02244$+$6117.
We confirmed that G\,134.2+0.8 associated with IRAS\,02245$+$6115 is 
excited by a deeply embedded young intermediate-mass star.
Although the W4 bubble and  \afg\  have previously interacted,  and induced mode of
star formation could be  responsible for the formation of  
SFO\,05 and IRAS\,02245$+$6115, the small velocity width, low rotational 
temperature of \nh3\ line, 
and low SFE
suggests that most stars in this region were formed by the quiescent mode
in the feedback-driven structure. 
Depending on the local environment,
both of the two modes of star formation, induced and quiescent,
coexist in scales of several pc or less.
It appears that star formation in this region has taken place
mainly without an external trigger, but accompanying stellar feedback environment.

\begin{ack}

We are grateful to the staff members of the Nobeyama Radio Observatory (NRO)
for observation support. We also thank John Bieging for providing Heinrich Hertz Telescope CO data cubes
and Takeshi Sakai for providing his NRO maps of the CO emission in FITS format. 
The research presented in this paper has used data from 
the Spitzer Space Telescope, which is operated by the Jet Propulsion Laboratory,
California Institute of Technology, under contract with NASA, and
the Canadian Galactic Plane Survey supported by the 
Natural Sciences and Engineering Research Council.
\end{ack}



\end{document}